\documentclass[reprint,twocolumn,floats, floatfix,superscriptaddress,amsmath,amssymb,aps]{revtex4-1}

\usepackage{graphicx}
\usepackage{dcolumn}
\usepackage{bm}
\usepackage{color}

\begin{document}


\title{Active instability and nonlinear dynamics of cell-cell junctions}

\author{Matej Krajnc}
\email{matej.krajnc@ijs.si}
\affiliation{Jo\v zef Stefan Institute, Jamova 39, SI-1000 Ljubljana, Slovenia}%

\author{Tomer Stern}
\affiliation{Department of Molecular Biology, Princeton University, Princeton, NJ, USA}%
\affiliation{Lewis-Sigler Institute for Integrative Genomics, Princeton University, Princeton, NJ 08544, USA}%

\author{Cl\'ement Zankoc}
\affiliation{Jo\v zef Stefan Institute, Jamova 39, SI-1000 Ljubljana, Slovenia}

\begin{abstract}
Active cell-junction remodeling is important for tissue morphogenesis, yet its underlying physics is not understood. We study a mechanical model that describes junctions as dynamic active force dipoles. Their instability can trigger cell intercalations by a critical collapse. Nonlinearities in tissue's elastic response can stabilize the collapse either by a limit cycle or condensation of junction lengths at cusps of the energy landscape. Furthermore, active junction networks undergo collective instability to drive active in-plane ordering or develop a limit cycle of collective oscillations, which extends over regions of the energy landscape corresponding to distinct network topologies.
\end{abstract}

\maketitle

\begin{figure}[b!]
    \includegraphics[]{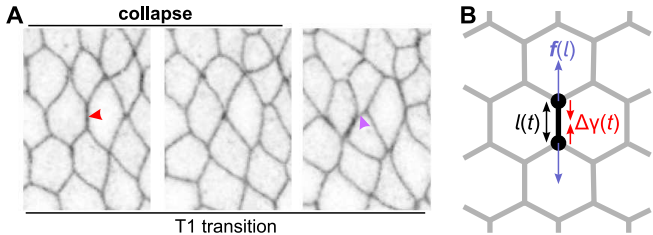}
    \caption{(A)~Snapshots of a T1 transition during GBE in {\it Drosophila}. Red and purple arrows point to a collapsing and to an extending junction, respectively. (B)~Schematic of the vertex model of an active junction. Junction length and active tension are denoted by $l(t)$ and $\Delta\gamma(t)$, respectively.  
    }
    \label{F1}
\end{figure}

{\it Introduction.---}Cell rearrangements are crucial for tissue deformations and flows during development, wound healing, and cancer~\cite{etournay15,tetley19,oswald17}. In confluent epithelia, cells rearrange through T1 transitions, where pairs of initially neighboring cells get separated by intercalation of adjacent cell pairs~({\color{blue}Fig.~\ref{F1}A} and Ref.~\cite{weaire84}). While in related passive materials, T1 transitions are induced by shear stresses or flows applied through system's boundaries~\cite{kraynik88,hohler05}, in tissues, they are usually driven locally by active contractions of individual cell-cell junctions. For instance, convergent extension in embryos occurs through planar-polarized active cell intercalations, driven by a relatively complex biomechanical machinery that includes multiple actomyosin structures~\cite{zallen04,bertet04,rauzi10}. In addition, active junctional noise can fluidize tissues and affects their in-plane organization~\cite{curran17,mongera18}.

Surprisingly, the biomechanics of active junctions are poorly understood theoretically~\cite{lenne21}. In particular, models of various morphogenetic events usually {\it impose} both spatial distribution and time course of active tensions and then compute cell deformations and flows using the relaxation dynamics~\cite{tetley19,zajac03,spahn13,rauzi13}. However, the dynamics of these active tensions play a crucial role in many biological contexts, e.g., the accelerated junctional contractions during {\it Drosophila}'s germband extension~(GBE)~\cite{tetley16,stern20} and a rapid buildup of force-generating myosin during wound closure~\cite{vedula15,kobb17}.

\begin{figure}[t!]
    \includegraphics[]{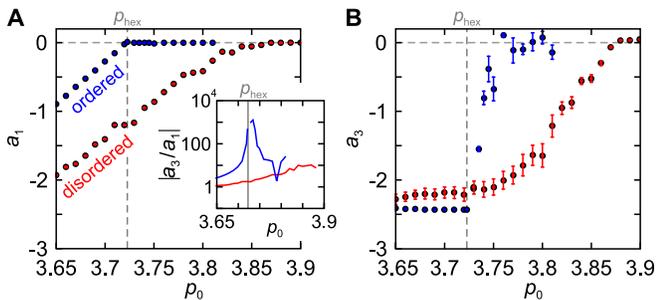}
    \caption{Linear and third-order elastic coefficients $a_1$ and $a_3$, resepectively, for $k_AA_0/k_P=100$. The elasticity of junctions in ordered tissues (blue points) is highly nonlinear around $p_0=p_{\rm hex}$, where the linear term vanishes. Junctions in disordered tissues (red points) are most nonlinear at $p_0\approx 3.86$. Inset: The ratio $\left | a_3/a_1\right |$ versus $p_0$.}
    \label{F2}
\end{figure}

It is also not understood how highly nonlinear elastic forces at the vicinity of the rigidity transition~\cite{staple10,bi15,sahu20} may affect the dynamics of junctions at the onset of cell movements. This can be studied using the Area- and perimeter-elasticity~(APE) vertex model, which represents the tissue by a planar tiling of polygonal cells, parametrized by the vertex positions $\boldsymbol r_i=\left (x_i,y_i\right )$~\cite{farhadifar07,fletcher14,alt17,barton17}. The potential energy of the system reads $W=\sum_{k}\left [k_A\left (A_k(t)-A_0\right )^2+k_P\left (P_k(t)-P_0\right )^2\right ]$~\cite{footnote1}, where the sum goes over all cells and describes cell-area elasticity ($A_k$ and $A_0$ being the actual and the preferred cell areas, respectively) and cell-perimeter elasticity ($P_k$ and $P_0$ being the actual and the preferred cell perimeters, respectively); $k_A$ and $k_P$ are the corresponding moduli. The nonlinear elastic response of junctions to a local force dipole at the vicinity of the rigidity transition shows in quasistatic force-extension curves, expanded around junction's rest length $l_0$: $f(l)=\sum_{i}a_i(l-l_0)^i$~\cite{footnoteX}, which are dominated by the third-order contribution~({\color{blue}Fig.~\ref{F2}} and {\color{blue}Supplemental Material, Fig.~S1~\cite{suppl}}). The elasticity is mainly controlled by the preferred cell-shape index $p_0=P_0/\sqrt{A_0}$ and depends on collective response to local active forces. The collective effects are best seen from distinct responses of ordered and disordered tissues: While ordered tissues respond highly nonlinearly around $p_0=p_{\rm hex}=2^{3/2}\cdot 3^{1/4}\approx 3.72442$~\cite{footnote2}, where $a_1$ vanishes, in disordered tissues this happens at $p_0\approx 3.86$~({\color{blue}Fig.~\ref{F2}}); overall, nonlinearities in disordered tissues are less dominant. Hysteresis effects may also be present, however, they turn out irrelevant for the interpretation of the results presented in this letter.

We study active cell-junction instability, driven by active force dipoles, generated in the junctional actomyosin. Our model describes critical junction collapse, which may provide a phenomenological explanation for accelerated contractions during morphogenesis. We find that nonlinearities in tissue's elastic response can stabilize the collapse, giving rise to a rich collection of active dynamics. For clarity, our analysis initially treats ordered tissues with a single active site. This is then generalized to disordered tissues and finally to networks of active junctions.

{\it The model.---}While cell-cell junctions actively contract during GBE, the concentration of junctional myosin locally increases, suggesting that molecular turnover cannot keep up with the rapid contractions. Interestingly, the contraction rate also increases, implying that the driving active tension may build up as well~({\color{blue}Supplemental Material, Fig.~S2~\cite{suppl}}). To describe this phenomenology, we follow a generic model of active contractile elements proposed by Dierkes {\it et al.}~\cite{dierkes14} and assume the active junctional tension $\Delta\gamma(t)$ proportional to the local concentration of myosin, defined as the number of motors per junction length, $c(t)=N(t)/l(t)$. In particular, $\Delta\gamma(t)=\alpha\left [c(t)-c_0\right ]$, where $c_0$ and $\alpha$ are the ambient myosin concentration and a constant proportionality factor, respectively. The total rate of change of the myosin concentration then reads $\dot c(t)=\dot N(t)/l(t)-N(t)\dot{l}(t)/l(t)^2$. Assuming motor-actin binding and undbinding at a rate $1/\tau_m$, described by $\dot{N}(t)=[-N(t)+c_0l(t)]/\tau_m$, yields a first-order equation for the active junctional tension:
\begin{equation}
    \label{eq:tensDyn}
    \Delta\dot\gamma(t)=-\frac{1}{\tau_m}\Delta\gamma(t)-\frac{\left [\gamma_0+\Delta\gamma(t)\right ]\dot l(t)}{l(t)}\>.
\end{equation}
Here, $\gamma_0=\alpha c_0$ is the tension at the ambient concentration of myosin motors. The first and the second term in Eq.~(\ref{eq:tensDyn}) describe relaxation due to myosin turnover and a mechanical feedback between junction contraction and force generation, respectively.

To describe tissue's mechanical response, we use the APE vertex model. Due to strong friction, described by an effective coefficient $\eta$, vertices follow the overdamped equation of motion:
\begin{align}
    \label{eq:vertexDyn}
    \eta\dot{\boldsymbol r}_i(t)=\boldsymbol F_i(t)=-\nabla_i W(\boldsymbol r_1,\boldsymbol r_2...)-\Delta\gamma(t)\nabla_il(t)\>.
\end{align}
Here, the first term describes restoring forces that drive the system towards the local minimum of $W$, whereas the second term describes local active forces. For simplicity, Eq.~(\ref{eq:vertexDyn}) assumes a single active junction with length $l(t)$, embedded in an otherwise passive tissue~({\color{blue}Fig.~\ref{F1}B}). The model is generalized to networks of active junctions in the penultimate section. To nondimensionalize Eqs.~(\ref{eq:tensDyn}) and (\ref{eq:vertexDyn}), we choose $\sqrt{A_0}$, $\tau_0=\eta/k_P$, and $k_P\sqrt{A_0}$ as the units of length, time, and tension, respectively. We study tissues with $324$ incompressible cells~($k_AA_0/k_P=100$), under periodic boundary conditions.
\begin{figure*}[t!]
    \includegraphics[]{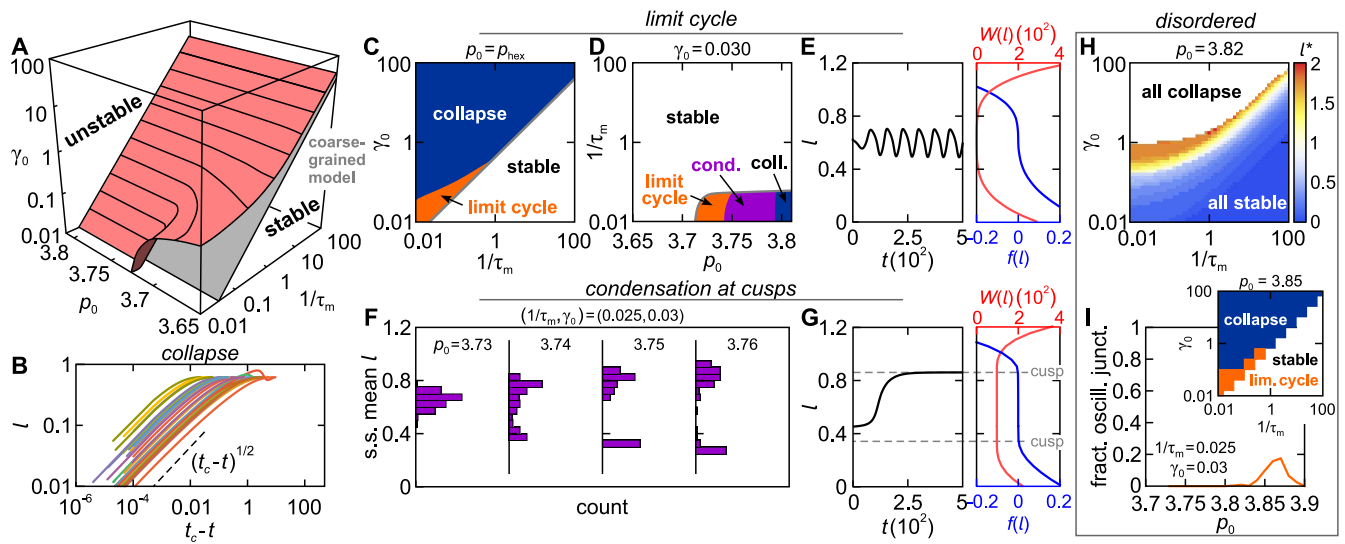}
    \caption{(A)~The critical bifurcation surface $\gamma^*(1/\tau_m,p_0)$ (red) and its coarse-grained approximation $\gamma_0^*=l_0/(2\tau_m)$ (gray). (B)~Junction length $l$ versus time $t_c-t$ for 25 randomly chosen collapse instances. Dashed line shows $l(t)\sim(t_c-t)^{1/2}$. (C,D)~Diagrams of junctional behaviors in $(1/\tau_m,\gamma_0)$ and $(p_0,1/\tau_m)$ planes at $p_0=p_{\rm hex}$ and $\gamma_0=0.03$, respectively. (E)~Junction length $l$ versus time $t$ at $(1/\tau_m,p_0,\gamma_0)=(0.19,p_{\rm hex},0.11)$~(left) and the corresponding local energy landscape, measured in a quasistatic limit. The landscape is dominated by third-order elasticity~(right). (F)~Distributions of steady-state mean junction lengths for $p_0=3.73-3.76$. (G)~Junction length $l$ versus time $t$ at $(1/\tau_m,p_0,\gamma_0)=(0.025,3.75,0.03)$~(left) and the corresponding local energy landscape, measured in a quasistatic limit. The landscape is dominated by higher-order nonlinear elasticity~(right). (H)~Critical junction rest length $l^*$, below which junctions collapse, in the $(1/\tau_m,\gamma_0)$ plane at $p_0=3.82$. (I)~Fraction of oscillating active junctions in disordered tissues versus $p_0$ at $(1/\tau_m,\gamma_0)=(0.025,0.03)$. Inset: Diagram of junctional behaviors in $(1/\tau_m,\gamma_0)$ plane at $p_0=3.85$.}
    \label{F3}
\end{figure*}

{\it Ordered tissues.---}The dynamical system described by Eqs.~(\ref{eq:tensDyn}) and (\ref{eq:vertexDyn}) has many fixed points, which describe equilibrium vertex configurations associated with distinct topologies of the cell-junction network. One such configuration is a regular honeycomb cell lattice, which is linearly stable in absence of activity, however only for $p_0<p_{\rm hex}$~\cite{staple10,farhadifar07}. Beyond this point, ordered tissues are no longer perfectly regular and may contain defects. Sine the results may depend on the position of these defects relative to the active site~({\color{blue}Supplemental Material, Fig.~S3~\cite{suppl}}), we make sure that when considering ordered tissues, the active junction is always surrounded by locally regular (honeycomb) cell tiling. 

Let us denote the fixed point of interest by $\boldsymbol\rho_0=\left (\boldsymbol r_1^{(0)},\boldsymbol r_2^{(0)}...,\boldsymbol r_{N_v}^{(0)},\Delta\gamma=0\right )$, where $\boldsymbol r_i^{(0)}$ are the equilibrium vertex positions. To study the linear stability of $\boldsymbol\rho_0$, we consider a linearized system $\delta \dot{\boldsymbol\rho}=\underline{\boldsymbol J}\delta\boldsymbol\rho$ for a perturbation $\delta\boldsymbol\rho=\left (\delta\boldsymbol r_1,\delta\boldsymbol r_2...,\delta\boldsymbol r_{N_v},\Delta\gamma\right )$, where $\underline{\boldsymbol J}$ describes the linearized system of Eqs.~(\ref{eq:tensDyn}) and (\ref{eq:vertexDyn}). We numerically diagonalize $\underline{\boldsymbol J}$ for different values of $1/\tau_m$, $p_0$, and $\gamma_0$ and identify the critical surface $\gamma_0^*=\gamma_0^*(1/\tau_m,p_0)$, where the stability switches~({\color{blue}Fig.~\ref{F3}A}). We find that the system undergoes a Hopf bifurcation either upon increasing $\gamma_0$ or decreasing $1/\tau_m$~({\color{blue}Supplemental Material, Fig.~S4~\cite{suppl}}).

Linear stability also depends on $p_0$, however, only for $p_0<p_{\rm hex}$. Beyond this critical point, the linear part of the elastic restoring force is zero~({\color{blue}Fig.~\ref{F2}A}) and so $p_0$, which governs tissue's response to the local activity, no longer plays a role. An approximate linear-stabillity condition for $p_0>p_{\rm hex}$ can be estimated by coarse-graining our model. Here, the length dynamics result from an interplay between the activity and friction: $\dot l(t)=-2\Delta\gamma(t)$, whereas $\Delta\dot\gamma(t)$ still obeys Eq.~(\ref{eq:tensDyn}). The system switches stability at $\gamma_0^*=l_0/(2\tau_m)$, where the nonzero eigenvalue of $\underline{\boldsymbol J}=\left ((0,-2),(0,-1/\tau_m+2\gamma_0/l_0)\right )$ changes sign~(gray plane in {\color{blue}Fig.~\ref{F3}A}).

To study the dynamics far from the fixed point, we next simulate the full nonlinear system~[Eqs.~(\ref{eq:tensDyn}) and (\ref{eq:vertexDyn})] in time upon a small random perturbation of all degrees of freedom from their fixed-point values. In the stable regime, the active junction either relaxes its length directly back to $l=l_0$ or it undergoes damped transient oscillations to the fixed point. Junctional noise~\cite{curran17,krajnc20} can amplify these oscillations, catching the system in a quasicycle around the stable fixed point~({\color{blue}Supplemental Material, Fig.~S4,  Movies M1 and M2~\cite{suppl}}, and Ref.~\cite{zankoc20}).

In the unstable regime, junctions undergo accelerated contractions, which are fundamentally different from a pure junction-length relaxation~({\color{blue}Supplemental Material, Fig.~S2 and Movie M3~\cite{suppl}}). By simulating the dynamics of 25 collapse instances, randomly selected from the collapse regime, we find that close to the point of vanishing length, the kinetics are independent of the parameter values~({\color{blue}Fig.~\ref{F3}B}). This is because the parameter-free second term in Eq.~(\ref{eq:tensDyn}) dominates as $l\to 0$. As a consequence, Eq.~(\ref{eq:tensDyn}) simplifies to $({\rm d}/{\rm d}t)(\Delta\gamma(t) l(t))=0$, implying $\Delta\gamma(t)\propto 1/l(t)$. In turn, in the coarse-grained approximation, $\dot l(t)=-2\Delta\gamma(t)$, thus yielding a governing equation for the dynamics of junction length close to $l=0$, which reads $\dot l(t)\propto l(t)^{-1}$. Its solution $l(t)\propto \left (t_c-t\right )^{1/2}$ agrees with the observed critical kinetics~({\color{blue}Fig.~\ref{F3}B}). These results show that our model may provide a phenomenological explanation for the accelerated junctional contractions during GBE, suggesting that cell intercalations may initiate through an active junctional instability. Note that the extent to which the predicted critical kinetics apply to biology is limited by the $l=0$ singularity~(Eq.~(\ref{eq:tensDyn}) and {\color{blue}Supplemental Material, Fig.~S2~\cite{suppl}}).

{\it Nonlinear effects.---}Next, we examine our system in the regime, where the elastic restoring force is highly nonlinear. We find that the third-order elastic term, which is dominant around $p_0\approx p_{\rm hex}$~({\color{blue}Fig.~\ref{F2}B}), can stabilize the collapse, giving rise to a limit cycle of junctional oscillations around a linearly unstable fixed point~({\color{blue}Fig.~\ref{F3}E} and {\color{blue}Supplemental Material, Movies M4 and M5~\cite{suppl}}). In particular, junctional oscillations are found for $3.71<p_0<3.75$ at sufficiently small values of $1/\tau_m$ and $\gamma_0$~({\color{blue}Fig.~\ref{F3}C and D}). The amplitude of length oscillations increases when approaching the transition to junction collapse, whereas their frequency increases with both $1/\tau_m$ and $\gamma_0$~({\color{blue}Supplemental Material, Fig.~S5~\cite{suppl}}).

Junctional oscillations only appear for values of $p_0$ up to $\approx 3.75$, where the third-order elastic term, needed to stabilize a limit cycle, vanishes and the energy landscape becomes locally flat~({\color{blue}Figs.~\ref{F3}D and \ref{F2}B}). However, we find that higher-order nonlinearities can also prevent junction collapse, yielding another type of dynamics around an unstable fixed point~({\color{blue}Fig.~\ref{F3}D}). To explore it, we simulate 100 trajectories at $(1/\tau_m,\gamma_0)=(0.025,0.030)$, each time randomly choosing a different active junction, and plot distributions of the steady-state mean lengths of these active junctions. While the unimodal distribution at $p_0=3.73$ implies that junction lengths are either stable or they oscillate around $l_0$, increasing $p_0$ turns this distribution into bimodal with the two peaks at $l>l_0$ and $l<l_0$~({\color{blue}Fig.~\ref{F3}F}). We find that these new stable lengths correspond to cusps in the energy landscape--regions where the landscape transitions from flat to nonflat~({\color{blue}Fig.~\ref{F3}G} and Ref.~\cite{sahu20}). Junctional noise can cause the system to switch between the two cusps, suggesting that they both act as linearly stable fixed points~({\color{blue}Supplemental Material, Fig.~S6 and Movies M6 and M7~\cite{suppl}}).

{\it Disordered tissues.---}Collapsing junctions in disordered tissues also follow the predicted critical kinetics~({\color{blue}Supplemental Material, Figs.~S3 and S7~\cite{suppl}}), however, their stabilities depend on rest lengths, which are distributed in disordered tissues. In fact, the stability condition can be recast in terms of the critical rest length $l^*$, below which junctions are expected to collapse~({\color{blue}Fig.~\ref{F3}H}). This bias provides a mechanism for an efficient active tissue ordering~(studied below).

The spatial variance in the network topology of disordered tissues leads to various types of active dynamics even for junctions within the same tissue. Nevertheless, simulations of 200 randomly chosen active sites shows that the different regions of the parameter space are dominated by same types of behaviors as in ordered tissues: While most active junctions collapse for small $1/\tau_m$- and big $\gamma_0$-values, they are mostly stable in the reverse case~(inset to {\color{blue}Fig.~\ref{F3}I}). The fraction of active junctions that develop a limit cycle, shows that oscillations appear in the vicinity of $p_0\approx 3.86$, where nonlinearities in the elastic response of disordered tissues dominate~({\color{blue}Fig.~\ref{F3}I} and {\color{blue}Fig.~\ref{F2}}).

{\it Networks of active junctions.---} The analysis of tissues with a single active site allowed us to characterize in detail the different types of active junctional dynamics both in ordered and disordered tissues. However, activity is usually patterned across tissues and so the interactions between multiple active sites are important as well. Rather than studying specific patterns of junctional activity, we next generalize our approach to the most generic case, where all junctions in the network are active and have equal mechanical properties. In this case, vertex $i$ obeys $\dot{\boldsymbol r}_i(t)=-\nabla_i W-\sum_{j}\Delta\gamma_j(t)\nabla_il_j(t)$, where the sum goes over all junctions $j$ that meet at vertex $i$; active tensions $\Delta\gamma_j$ evolve in time according to~Eq.~(\ref{eq:tensDyn}).

\begin{figure}[t!]
    \includegraphics[]{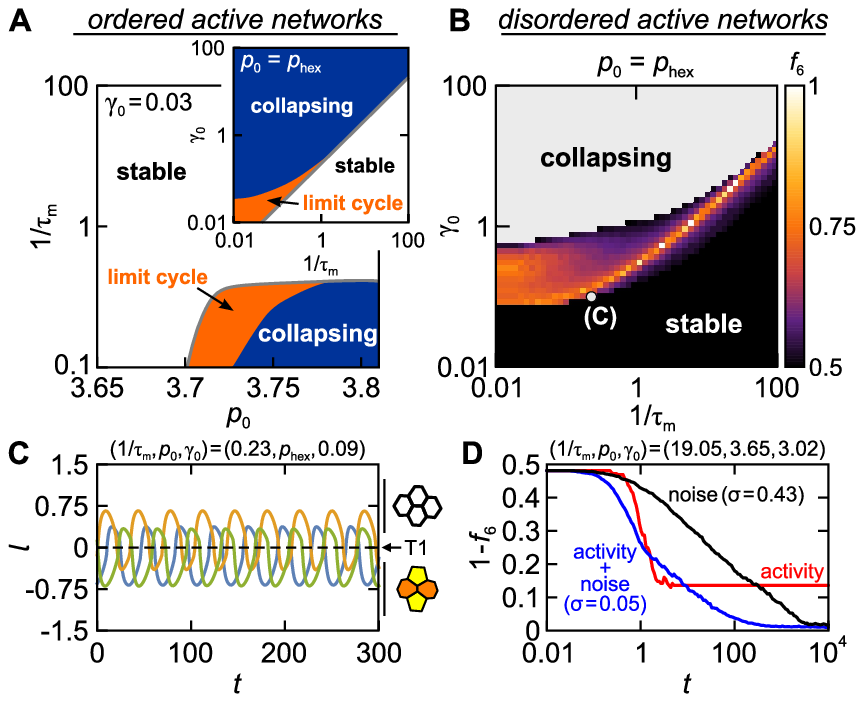}
    \caption{(A)~Junctional behaviors in ordered active networks in the $(p_0,1/\tau_m)$ plane at $\gamma_0=0.03$ and in the $(1/\tau_m,\gamma_0)$ plane at $p_0=p_{\rm hex}$ (inset). (B)~Steady-state fraction of hexagonal cells $f_6$ after active ordering in the $(1/\tau_m,\gamma_0)$ plane at $p_0=p_{\rm hex}$. (C)~Junction length $l$ vs. time $t$ for three junctions that undergo T1 transitions during collective junctional oscillations at $(1/\tau_m,p_0,\gamma_0)=(0.23,p_{\rm hex},0.09)$~[denoted by (C) in panel B]. (D)~Fraction of non-hexagonal cells $1-f_6$ as a function of time $t$ for tissue ordering at $(1/\tau_m,p_0,\gamma_0)=(19.05,3.65,3.02)$, driven by activity (red), pure noise (black), and activity+noise (blue).}
    \label{F4}
\end{figure}

The diagram of junctional behaviors of ordered active networks is similar to that of a single active junction: The limit cycle, describing collective junctional oscillations, appears for $p_0\approx p_{\rm hex}$ in the regime of small $1/\tau_m$ and $\gamma_0$~({\color{blue}Fig.~\ref{F4}A}). In turn, the "collapsing" regime describes tissues, in which all junctions can be viewed as unstable and as a result, cell intercalations occur throughout the tissue, which thus becomes highly disordered.

In contrast, disordered active networks undergo active ordering in the regime, where the critical rest length for individual-junction collapse is neither 0 nor too big~({\color{blue}Fig.~\ref{F4}B} and {\color{blue}Fig.~\ref{F3}H}). Here, short junctions are unstable and undergo T1 transitions, which persist as long as the tissue is sufficiently disordered to contain short junctions. In fact, this mechanism naturally implements a greedy-type optimization algorithm~\cite{greedyBook}. Indeed, the global tissue ordering is driven by local optimization steps, which transform short (unstable) junctions into long (stable) ones through T1 transitions~({\color{blue}Supplemental Material, Fig.~S8~\cite{suppl}}). 

Around $p_0\approx p_{\rm hex}$, tissues may establish collective oscillations as they transition to an ordered state. Since the steady-state ordered configuration is essentially never defect-free, these oscillations can be spatially quite heterogeneous. Interestingly, we even find cases, where the limit cycle includes periodic T1 transitions~({\color{blue}Fig.~\ref{F4}C} and {\color{blue}Supplemental Material, Movie M8~\cite{suppl}}). In these cases, the trajectory cycles around fixed points corresponding to distinct network topologies. 

Tissue ordering is important for maintaining homeostatic conditions and has been observed in {\it Drosophila}'s pupal notum, where it is driven by active tension fluctuations~\cite{curran17,sugimura13}. In our model, these fluctuations may enter $\Delta\dot\gamma_j$ by a term $\sqrt{2\sigma^2/\tau_m}\>\xi_j(t)$, describing the white noise with long-time variance $\sigma^2$; $\langle\xi_j(t)\rangle=0$, $\langle\xi_j(t)\xi_k(t')\rangle=\delta_{jk}\delta(t-t')$. We find that the mechanical feedback described in our model~[Eq.~(\ref{eq:tensDyn})], can speed up noise-driven ordering by as much as one order of magnitude~({\color{blue}Supplemental Material, Fig.~S9~\cite{suppl}}). In most severe cases, disordered networks can clear out all defects and become perfectly ordered on a time scale of $\sim 100\tau_0$~({\color{blue}Fig.~\ref{F4}D} and {\color{blue}Supplemental Material, Movies M9 and M10~\cite{suppl}}). 

{\it Discussion.---}Our model suggests that in-plane tissue remodelling during morphogenesis may be locally triggered by an active junctional instability, yielding an accelerated junctional collapse. In turn, nonlinearities in tissue's elastic response appearing at the vicinity of the rigidity transition may stabilize the collapse, giving rise to various types of junctional movements. Junctional stability also depends on the junction's rest length, which enables active tissue ordering, relevant for maintaining homeostatic conditions in developing tissues. Our work opens new avenues of possibilities for future studies. In particular, the extent to which the predicted nonlinear junctional behaviors appear {\it in vivo} under different biological conditions may be probed by means of experimental techniques that rely on high spatial and temporal resolutions, e.g., the spinning-disc confocal imaging. Additionally, the relative role of the active junctional instability may be assessed using optogenetic manipulations~\cite{staddon19}. Our model may also be generalized by including additional biochemical mechanisms such as the tissue-scale patterning and contributions of non-junctional actomyosin structures~\cite{kong17,rauzi20}. Finally, the active junctional dynamics may be affected by additional sources of energy dissipation, related to changes of cell's geometric elements, which may be included in our model through a friction matrix~\cite{alt17}. Not only these additional force components would explicitly affect vertex movements, they could also considerably affect the generation of active forces through the mechanical feedback~[Eq.~(\ref{eq:tensDyn})].
\begin{acknowledgments}
We thank Simon \v Copar for the suggestion to look at the critical dynamics of junction collapse, Ricard Alert, Eric Wieschaus, Stas Shvartsman, and other members of the Princeton's Gastrulation club for fruitful discussions, and Jan Rozman, and Primo\v z Ziherl for critical reading of the manuscript. We acknowledge the financial support from the Slovenian Research Agency (research project No. Z1-1851 and research core funding No. P1-0055).
\end{acknowledgments}


\begin{thebibliography}{99}

\bibitem{etournay15}
    R.~Etournay, M.~Popovi{\'c}, M.~Merkel, A.~Nandi, C.~Blasse, B.~Aigouy, H.~Brandl, G.~Myers, G.~Salbreux, F.~J{\"u}licher, S.~Eaton, and H.~McNeill, eLife {\bf 4}, e07090 (2015).

\bibitem{tetley19}
	R.~J.~Tetley, M.~F.~Staddon, D.~Heller, A.~Hoppe, S.~Banerjee, and Y.~Mao, Nat. Phys. {\bf 15}, 1195 (2019).
	
\bibitem{oswald17}
	L. Oswald, S. Grosser, D.~M.~Smith, and J.~A.~K\"as, J. Phys. D: Appl. Phys. {\bf 50}, 483001 (2017).
	
\bibitem{weaire84}
	D. Weaire and N. Rivier, Contemp. Phys. {\bf 25}, 59 (1984).
	
\bibitem{kraynik88}
	A.~M.~Kraynik, Annu. Rev. Fluid Mech. {\bf 20}, 325 (1988).
	
\bibitem{hohler05}
	R.~H\"ohler and S.Cohen-Addad, J. Phys. Condens. Matter {\bf 17}, R1041 (2005).
	
\bibitem{zallen04}
	J.~A.~Zallen and E.~F.~Wieschaus, Dev. Cell {\bf 6}, 343 (2004).
	
\bibitem{bertet04}
	C.~Bertet, L.~Sulak, and T.~Lecuit, Nature {\bf 429}, 667 (2004).
	
\bibitem{rauzi10}
	M.~Rauzi, P.-F.~Lenne, and T.~Lecuit, Nature {\bf 468}, 1110 (2010).
	
\bibitem{curran17}
	S.~Curran, C.~Strandkvist, J.~Bathmann, M.~de~Gennes, A.~Kabla, G.~Salbreux, and B.~Baum, Dev. Cell {\bf 43}, 480 (2017).

\bibitem{mongera18}
  	A.~Mongera, P.~Rowghanian, H.~J.~Gustafson, E.~Shelton, D.~Kealhofer, E.~K.~Carn, F.~Serwane, A.~A.~Lucio, J.~Giammona, and O. Camp\`as, Nature {\bf 561}, 401 (2018).

\bibitem{lenne21}
  	P.-F.~Lenne, J.-F.~Rupprecht, and V.~Viasnoff, Dev. Cell {\bf 56}, 202 (2021).
	
\bibitem{zajac03}
	M.~Zajac, G.~L.~Jones, and J.~A.~Glazier, J. Theor. Biol. {\bf 222}, 247 (2003).
	
\bibitem{spahn13}
	P.~Spahn and R.~Reuter, PLOS One {\bf 8}, e75051 (2013).
	
\bibitem{rauzi13}
	M.~Rauzi, A.~Ho\v cevar~Brezav\v s\v cek, P.~Ziherl, and M.~Leptin, Biophys. J. {\bf 105}, 3 (2013).

\bibitem{tetley16}
	R.~J.~Tetley, G.~B.~Blanchard, A.~G.~Fletcher, R.~J.~Adams, and B.~Sanson, eLife {\bf 5}, e12094 (2016).
	
\bibitem{stern20}
	T.~Stern, S.~Y.~Shvartsman, and E.~F.~Wieschaus, PLOS Comput. Biol. {\bf 16}, 1 (2020).
	
\bibitem{vedula15}
	S.~R.~K.~Vedula, G.~Peyret, I.~Cheddadi, T.~Chen, A.~Brugues, H.~Hirata, H.Lopez-Menendez, Y.~Toyama, L.~N.~de Almeida, X.~Trepat, C.~T.~Lim, and B.~Ladoux, Nat. Commun. {\bf 6}, 6111 (2015).
	
\bibitem{kobb17}
	A.~B.~Kobb, T. Zulueta-Coarasa, and R. Fernandez-Gonzalez, J. Cell Sci. {\bf 130}, 689 (2017).
	
\bibitem{staple10}
	D.~B.~Staple, R.~Farhadifar, J.-C.~R\"oper, B.~Aigouy, S.~Eaton, and F.~J\"ulicher, Eur. Phys. J. E {\bf 33}, 117 (2010).
	
\bibitem{bi15}
	D.~Bi, J.~Lopez, J.~Schwarz, and M.~L.~Manning, Nat. Phys. {\bf 11}, 1074 (2015).
	
\bibitem{sahu20}
	P.~Sahu, J.~Kang, G.~Erdemci-Tandogan, and M.~L.~Manning, Soft Matter {\bf 16}, 1850 (2020).
	
\bibitem{farhadifar07}
	R.~Farhadifar, J.-C.~R\"oper, B.~Aigouy, S.~Eaton, and F.~J\"ulicher, Curr. Biol. {\bf 17}, 2095 (2007).
	
\bibitem{fletcher14}
	A.~Fletcher, M.~Osterfield, R.~Baker, and S.~Y.~Shvartsman, Biophys. J. {\rm 106}, 2291 (2014).
	
\bibitem{alt17}
	S.~Alt, P.~Ganguly, and G.~Salbreux, Philos. Trans. Royal Soc. B {\bf 372}, 20150520 (2017).
	
\bibitem{barton17}
	D.~L.~Barton, S.~Henkes, C.~J.~Weijer, and R.~Sknepnek, PLoS Comp. Biol. {\bf 13}, e1005569 (2017).
	
\bibitem{footnote1}
    The potential energy of the tissue in its full form reads $W=\sum_kk_A\left(A_k-A_0\right )^2+\sum_kk_PP_k^2-\sum_{ij}\alpha l_{ij}+\sum_{ij}\gamma_0l_{ij}$. Here $k_A$ and $k_P$ are the moduli of the cell- and perimeter elasticity, respectively, $\alpha$ is the adhesion strength, and $\gamma_0$ is the tension in the junctional actomyosin. By defining the preferred perimeter as $P_0=(\alpha-\gamma_0)/(4k_P)$, we can write $W=\sum_{k}\left [k_A(A_k-A_0)^2+k_P(P_k-P_0)^2\right ]$.
    
\bibitem{footnoteX}
	Force $f(l)$ is a quasistatic elastic restoring force, which acts on a compressed/stretched junction. It is measured by quasistatically varying junction length $l$ and computing $f(l)=-\nabla_i W\cdot\nabla_i l-\nabla_j W\cdot\nabla_j l$; here $i$ and $j$ are ids of vertices that constitute the junction in question.
    
\bibitem{suppl}
	See Supplemental Material at http://link.aps.org/...
    
\bibitem{footnote2}
    The value $p_{\rm hex}\approx 3.72442$ corresponds to the perimeter of a regular hexagon with unit surface area.
	
\bibitem{dierkes14}
	K.~Dierkes, A.~Sumi, J.~Solon, and G.~Salbreux, Phys. Rev. Lett {\bf 113}, 148102 (2014).
	
\bibitem{krajnc20}
	M.~Krajnc, Soft Matter {\bf 16}, 3209 (2020).
	
\bibitem{zankoc20}
	C.~Zankoc and M. Krajnc, Biophys. J. {\bf 119} 1706 (2020).
	
	
\bibitem{greedyBook}
	T.~Roughgarden, Algorithms Illuminated (Part 3): Greedy Algorithms and Dynamic Programming (Soundlikeyourself Publishing, New York, 2019).
	
\bibitem{sugimura13}
	K.~Sugimura and S.~Ishihara, Development {\rm 140}, 4091 (2013).
	
\bibitem{staddon19}
	M.~F.~Staddon, K.~E.~Cavanaugh, E.~M.~Munro, M.~L.~Gardel, and S.~Banerjee, Biophys. J. {\bf 117}, 1739 (2019).
	
\bibitem{kong17}
	D.~Kong, F.~Wolf, and J.~Grosshans, Mech. Dev. {\bf 144}, 11 (2017).
	
\bibitem{rauzi20}
	M.~Rauzi, Philos. Trans. R. Soc. B {\bf 375}, 20190552 (2020).
	
	
	

	
	

	
\end{thebibliography}

\end{document}




\title{Active instability and nonlinear dynamics of cell-cell junctions:\\Supplemental material}

\author{Matej Krajnc}
\affiliation{Jo\v zef Stefan Institute, Jamova 39, SI-1000 Ljubljana, Slovenia}%

\author{Tomer Stern}
\affiliation{Department of Molecular Biology, Princeton University, Princeton, NJ, USA}%
\affiliation{Lewis-Sigler Institute for Integrative Genomics, Princeton University, Princeton, NJ 08544, USA}%

\author{Cl\'ement Zankoc}
\affiliation{Jo\v zef Stefan Institute, Jamova 39, SI-1000 Ljubljana, Slovenia}


\maketitle

\onecolumngrid

\section{Details of implementation}
A tissue is described by a planar tiling of polygonal cells, parametrized by the positions of vertices $\boldsymbol r_{i}=(x_i,y_i)$. Due to friction with the environment, described by the coefficient $\eta$, vertices obey the overdamped equation of motion, which reads
\begin{equation}
	\eta\dot{\boldsymbol r}_i(t)=\boldsymbol F_i=\boldsymbol F^{(R)}_i+\boldsymbol F^{(A)}_i\>,
\end{equation}
where $\boldsymbol F_i^{(R)}$ and $\boldsymbol F^{(A)}_i$ are the restoring and active forces, respectively. 

{\bf Restoring forces.} The restoring forces $\boldsymbol F_i^{(R)}=-\nabla_i W(\boldsymbol r_1,\boldsymbol r_2...)$ drive the system towards the local minimum of the potential energy $W=\sum_{k}\left [k_A\left (A_k(t)-A_0\right )^2+k_p\left (P_k(t)-P_0\right )^2\right ]$. Here $A_k$ and $A_0$ are the actual and the preferred surface area of cell $k$, respectively, whereas $P_k$ and $P_0$ are the actual and the preferred perimeter of cell $k$, respectively; $k_A$ and $k_P$ are the corresponding moduli. Choosing $\sqrt{A_0}$, $\eta/k_p$, and $k_p\sqrt{A_0}$ as the units of length, time, and tension, respectively, yields a dimensionless restoring force, which reads
\begin{equation}
	\boldsymbol F_i^{(R)}=-2\sum_{k}\left [\left (A_k-1\right )\nabla_i A_k+\left (p_k-p_0\right )\nabla_ip_k\right ]\>.
\end{equation}
The dimensionless cell perimeter $p_k=\sum_{\mu\in{\rm cell}\>k} \left |{\boldsymbol r}_{\mu+1}-{\boldsymbol r}_{\mu}\right |$, whereas the dimensionless cell area $A_k=(1/2)\sum_{\mu\in{\rm cell}\>k}\left ({\boldsymbol r}_{\mu+1}\times{\boldsymbol r}_\mu\right )\cdot\hat{\boldsymbol e}_z$; the unit vector $\hat{\boldsymbol e}_z$ points in the direction perpendicular to the polygons's surface. The sums in both $p_k$ and $A_k$ go over all vertices of cell $k$ in the anti-clockwise direction.

{\bf Active forces.} Activity is modelled by active force dipoles, i.e., pairs of equal and opposite attractive forces between vertices belonging to active junctions. In case of a single active site, the active force on vertex $i$ reads $\boldsymbol F^{(A)}_i=-\Delta\gamma\nabla_il$, where $\Delta\gamma$ and $l$ are dimensionless active tension and length of the active junction, respectively. While $\Delta\gamma$ is the magnitude of $\boldsymbol F^{(A)}_i$, $-\nabla_il$ specifies its directionality. Indeed, $-\nabla_il=-\nabla_i\sqrt{(x_i-x_j)^2+(y_i-y_j)^2}=({\bf r}_j-{\bf r}_i)/\left |{\bf r}_j-{\bf r}_i\right |$, where $j$ is the id of the other vertex of the active junction. The force $\boldsymbol F^{(A)}_i$ is nonzero only for the two vertices corresponding to the active junction; for all other vertices $i$, $\nabla_i l=0$. The junctional tension at the active site obeys $\Delta\dot\gamma(t)=-(1/\tau_m)\Delta\gamma(t)-\left [\gamma_0+\Delta\gamma(t)\right ]\dot l(t)/l(t)$, where $1/\tau_m$ and $\gamma_0$ are the myosin turnover rate and tension at the ambient concentration of myosin motors, respectively. The rate of active-junction contraction can be expressed in terms of forces acting on the two vertices of the active junction, here denoted by indices 1 and 2: $\dot l=({\boldsymbol r}_2-{\boldsymbol r}_1)\cdot({\boldsymbol F}_2-{\boldsymbol F}_1)/l$, whereas $l=\sqrt{({\boldsymbol r}_2-{\boldsymbol r}_1)\cdot ({\boldsymbol r}_2-{\boldsymbol r}_1)}$. In case of active junctional networks, in which all junctions are considered active, the active force on vertex $i$ is nonzero for all vertices and reads $\boldsymbol F^{(A)}_i=-\sum_{j}\Delta\gamma_j(t)\nabla_il_j(t)$, where the sum goes over all junctions $j$ that meet at vertex $i$. The active junctional tensions $\Delta\gamma_j$ again obey $\Delta\dot\gamma_j(t)=-(1/\tau_m)\Delta\gamma_j(t)-\left [\gamma_0+\Delta\gamma_j(t)\right ]\dot l_j(t)/l_j(t)$.

The system of differential equations for $\boldsymbol r_i(t)$ and $\Delta\gamma(t)$ (single active site) or $\Delta\gamma_j(t)$ (active junctional networks) is solved using an Explicit Euler integration scheme; the upper limit for the time step used is $10^{-3}$; the time step is adapted according to the maximal vertex displacement, which is never allowed to exceed $\left |\Delta {\boldsymbol r}\right |=10^{-3}$.

In simulations that include T1 transitions (i.e., active junction networks), a T1 transition is initiated as soon as the junction length drops below $0.01$. The length of the newly created junction after the edge swap is set to $10^{-3}$, whereas the tension $\Delta\gamma$ is set to 0.
\section{Preparation of ordered and disordered tissue samples}
To prepare ordered and disordered tissue samples, we use a pure Ornstein-Uhlenbeck scheme to describe active junctional tensions. In particular, the dynamics of tension fluctuations on junction $j$ are given by
\begin{equation}
   \dot{\Delta\gamma_j}=-\frac{1}{\tau_m}\Delta\gamma_j+\sqrt{\frac{2\sigma^2}{\tau_m}}\>\xi_j\>,
\end{equation}
where the second term describes the white noise with long-time variance $\sigma^2$; $\langle\xi_j(t)\rangle=0$, $\langle\xi_j(t)\xi_k(t')\rangle=\delta_{jk}\delta(t-t')$.

We start each simulation with a regular honeycomb lattice, which we fluidize by running simulations at $\sigma=0.5$ and $\tau_m=1$ for $\Delta t=3000$, such that the tissue becomes highly disordered. Next, we quench the system instantaneously by setting $\sigma$ to a finite value $\sigma\leq 0.5$ and simulate the model for another $\Delta t=7000$. Finally, we set $\sigma$ to 0 and run the simulation for another $\Delta t=1000$ at a fixed topology of the cell-junction network so as to reach the local minimum of the potential energy.

Depending on the value of $\sigma$, the tissue reaches different degrees of disorder (i.e., fraction of hexagonal cells) by the end of simulation. For ordered tissues samples, we use those that were generated at $\sigma$-values that yield most ordered tissues with the fraction of hexagonal cells close to 1~(Fig.~S3A), whereas for disordered tissue samples, we use those that were generated at $\sigma=0$. In the latter case, tension fluctuations are no longer present after the initial disordering phase. As a result, the second phase yields only partial tissue ordering due to T1 transitions that occur spontaneously as the system dissipates energy through friction. These T1 transitions stop when there are no more vanishingly short junctions. It turns out that the fraction of hexagonal cells in the thus generated disordered tissue samples is always around $f_6=0.5$~(Fig.~S3B).
\section{Experiments}
We used a confocal microscope (Leica SP5) to acquire $X\times Y\times Z\times t$ stacks of a gastrulating fly embryo carrying a sqh-Cherry and E-cad EGFP markers with spatial resolution of $0.3079\>\mu {\rm m}\times 0.3079\>\mu {\rm m}\times 0.5\>\mu {\rm m}$, and temporal resolution of $32\>{\rm s}$ for a total period of $30\>{\rm min}$. Data segmentation was done by applying a max projection on each $XYZ$ stack, followed by applying cell segmentation using the publicly available image segmentation software \texttt{cell-pose}. Then, cells were tracked by applying non-rigid deformation between pairs of consecutive time points using Demon’s algorithm and identifying correspondences based on spatial overlap. Junction intensity was defined as the average pixel intensity in all pixels residing at the interface between the two neighboring cells (values are reported in arbitrary units), and junction length as the Euclidean distance between the two farthest pixels in the junction. To generate graphs shown in Fig.~S2A and B, we isolated interfaces that persist for at least $15\>{\rm min}$ and result in complete elimination (i.e. $l\to 0$), which yield a set of 110 sequences. Finally, we aligned the sequences temporally according to the time of elimination.
\newpage
\section{Supplemental Figures}
	\begin{figure}[h]
		\includegraphics{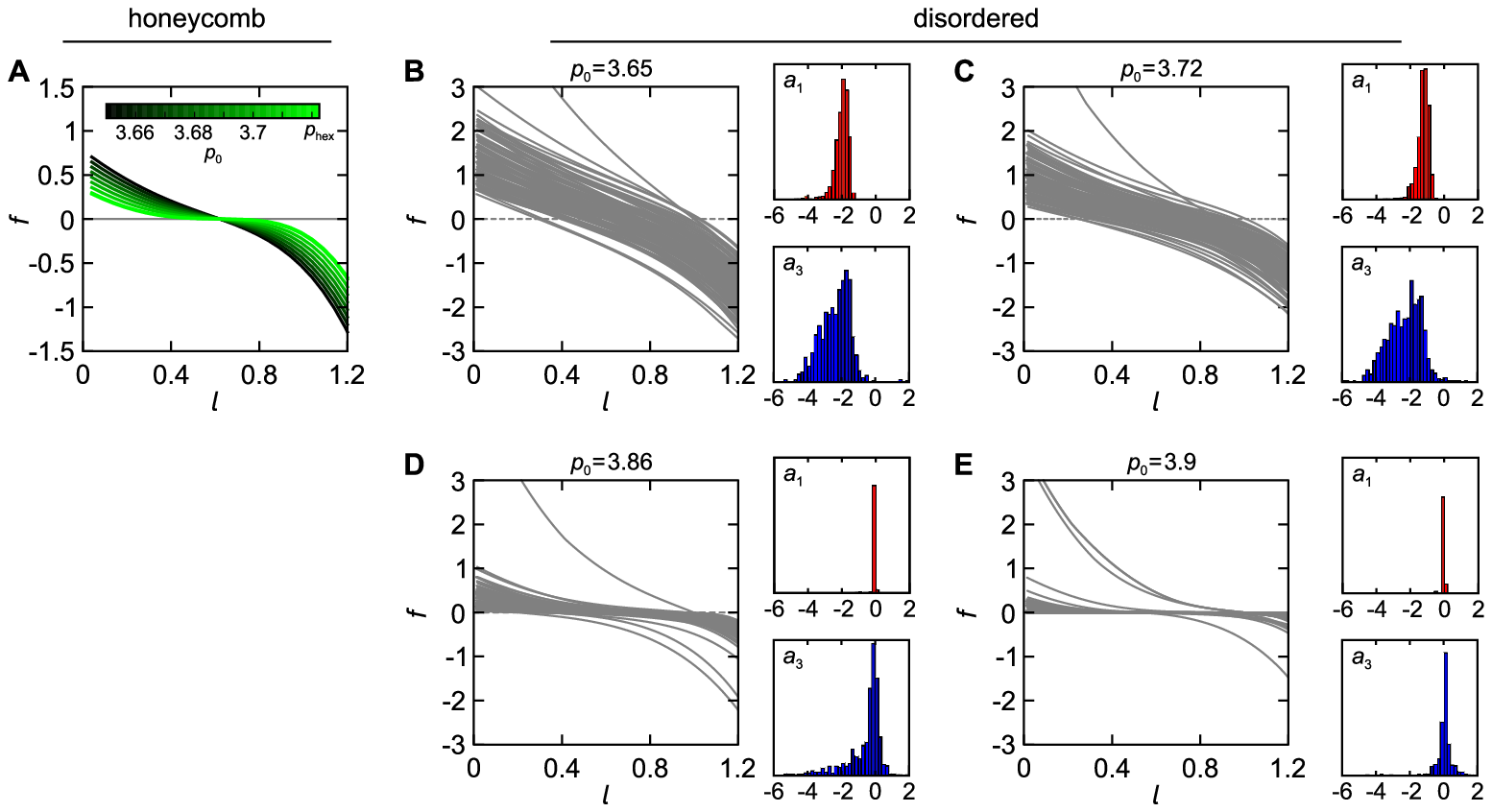}
		\caption{(A)~Force-length curves for a junction in a honeycomb cell lattice. The black-to-green color scheme encodes the value of $p_0$. As $p_0$ approaches $p_{\rm hex}$, $f(l)$ become highly nonlinear around the rest length $l=l_0$. (B-E)~In disordered tissues samples, $f(l)$ were measured separately for all 972 junctions. Gray curves show $f(l)$ of randomly chosen $N=100$ junctions at $p_0=3.65,\>3.72,\>3.86$ and $3.9$ (panels B-E, respectively). At each $p_0$ value, all 972 curves were fitted by a polynomial $f(l)=\sum_{i}a_i(l-l_0)^i$ and first- and third-order coefficients $a_1$ and $a_3$, respectively, were extracted. The distributions of $a_1$ and $a_2$ for all 972 junctions are plotted next to each $f(l)$ plot (red and blue columns, respectively). }
	\end{figure}

	\begin{figure}[h]
		\centering
		\includegraphics{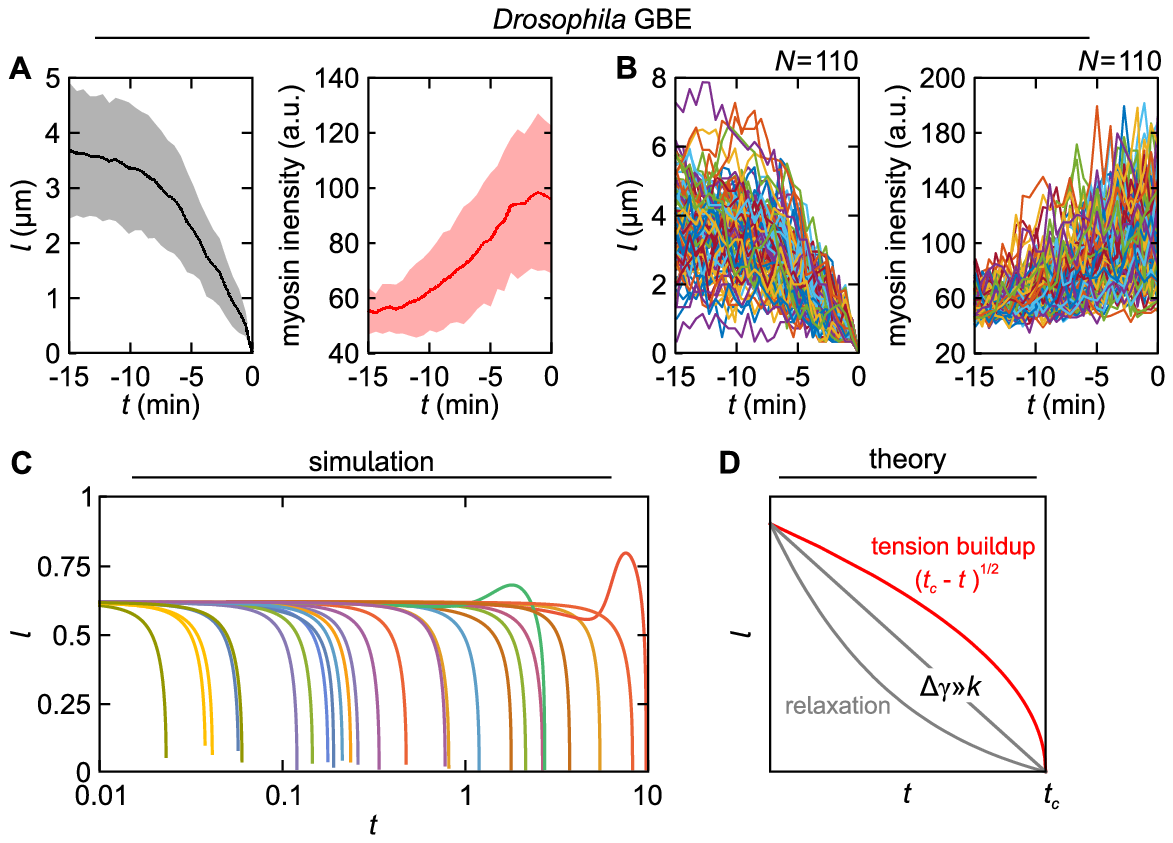}
		\caption{(A)~Junction length $l$ versus time $t$~(left) and the corresponding junctional-myosin intensity versus time $t$~(right) showing the average junction-collapse dynamics during {\it Drosophila}'s germband extension~(GBE). The solid curves and the shaded regions show the mean and the standard deviation over $N=110$ analyzed junctions, respectively. The raw data are shown in panel B. (C)~Junction length $l$ versus time $t$ for 25 collapse instances, randomly chosen from the unstable region of the parameter space~(Fig.~3A). (D)~Junction contraction during T1 transition can be studied in a coarse-grained fashion by an overdamped equation of motion for junciton length, which in dimensionless form reads $\dot l=-2\Delta\gamma-k(l-l_0)$, and describes the competition between the active force $\Delta\gamma$ and the elastic restoring force with the associated spring constant $k$. This equation yields a relaxation-like kinetics of junction contraction (gray curves), which can be at most linear in the limit, where the activity dominates ($\Delta\gamma\gg k$). In contrast, our model, which assumes $\Delta\gamma=\Delta\gamma(t)$, predicts an accelerated junctional collapse, triggered by a mechanical instability. Close to the point of vanishing junction length, the collapse follows the critical $\sim(t_c-t)^{1/2}$ kinetics (red curve).
		}\vspace{0.5cm}
	\end{figure}
	
	\begin{figure}[h!]
		\centering
		\includegraphics{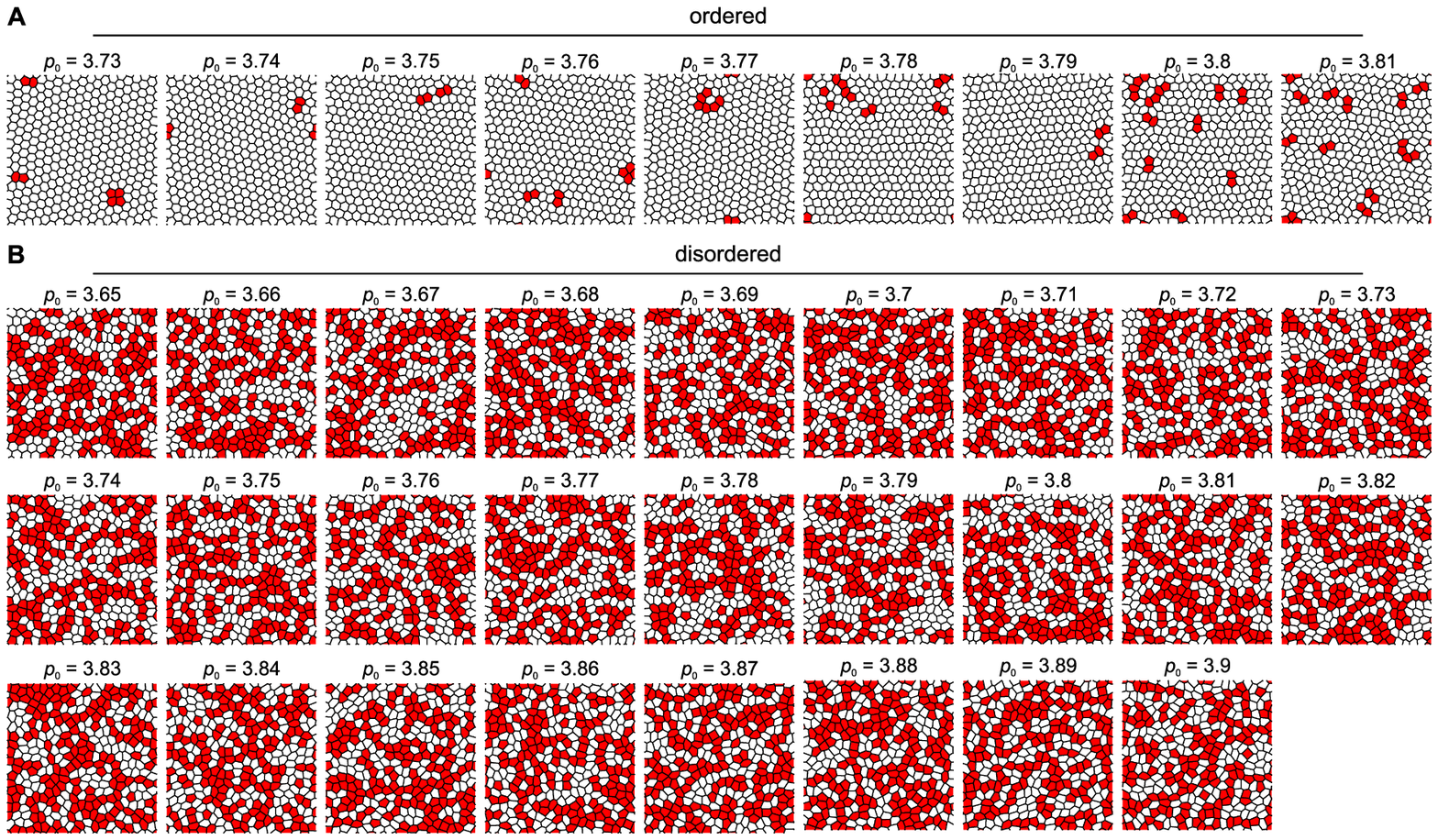}
		\caption{Ordered (panel A) and disordered (panel B) tissue samples generated to study active junctional dynamics. The fraction of hexagonal cells (shown in white) is close to 1 in ordered tissue samples and around 0.5 in disordered tissue samples.} 
	\end{figure}
	\begin{figure}[h]
		\centering
		\includegraphics{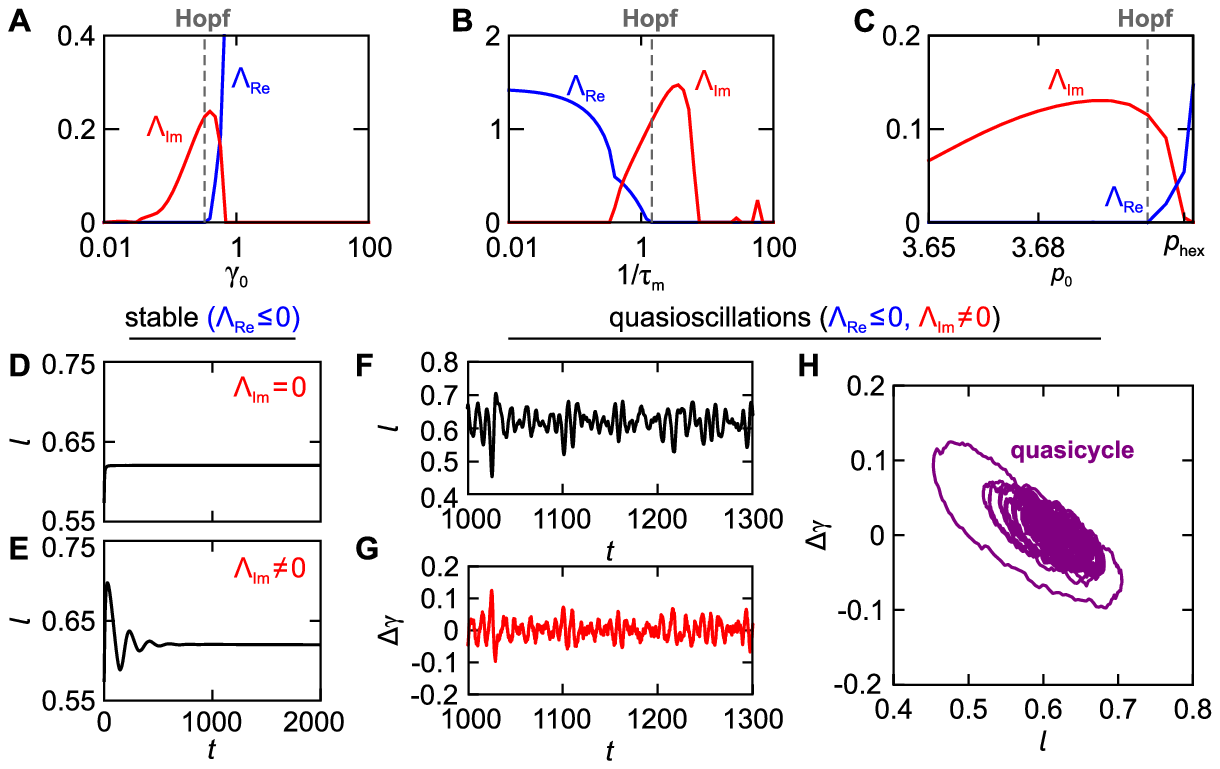}
		\caption{(A-C)~Maximal real and imaginary parts of the eigenvalues of the Jacobian matrix $\underline{\boldsymbol J}$, $\Lambda_{\rm Re}$ and $\Lambda_{\rm Im}$, respectively, versus $\gamma_0$~(panel A), $1/\tau_m$~(panel B), and $p_0$~(panel C). In panel A: $1/\tau_m=0.11$ and $p_0=3.65$; In panel B: $\gamma_0=1$ and $p_0=3.65$; In panel C: $1/\tau_m=\gamma_0=0.11$. (D,~E)~Upon a perturbation, junction length relaxes directly back to the fixed point if $\Lambda_{\rm Im}=0$~(panel D), and undergoes damped transient oscillations to the fixed point if $\Lambda_{\rm Im}\neq0$~(panel E). Panel D: $p_0=3.7$, $1/\tau_m=0.01$, and $\gamma_0=0.09$; Panel E: $p_0=3.7$, $1/\tau_m=15.85$, and $\gamma_0=0.09$. (F,~G)~Junction length $l$ (panel F) and active tension $\Delta\gamma$ (panel G) versus time $t$ at $p_0=3.65$ and $1/\tau_m=\gamma_0=0.575$. (H)~The dynamics from panels F and G shown in the $(l,\Delta\gamma)$ plane. The trajectory is called quasicycle and corresponds to transient oscillations to the stable fixed point, sustained and amplified by junctional noise. Noise is described by an additional term in the equation for active tension [Eq.~(1) of the main text]: $\Delta\dot{\gamma}_{\rm noise}=\sqrt{2\sigma^2/\tau_m}\>\xi(t)$, which describes the white noise with long-time variance $\sigma^2$ and $\langle\xi(t)\rangle=0$, $\langle\xi(t)\xi(t')\rangle=\delta(t-t')$. The results shown in panels F-H correspond to $\sigma=0.01$.}\vspace{0.5cm}
	\end{figure}
	
	\begin{figure}[h]
		\centering
		\includegraphics{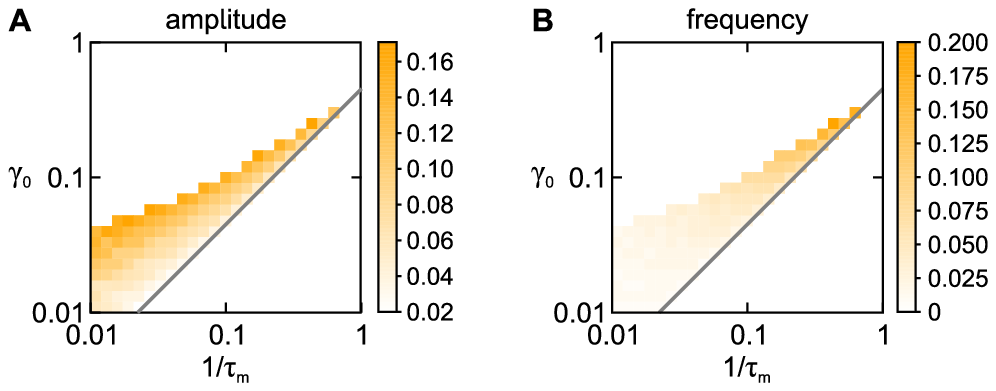}
		\caption{Amplitude~(panel A) and frequency~(panel B) of junction-length oscillations for the case of a single active junction in ordered tissues in the $(1/\tau_m,\gamma_0)$-plane at $p_0=p_{\rm hex}$.}\vspace{0.5cm}
	\end{figure}
	
	\begin{figure}[h]
		\centering
		\includegraphics{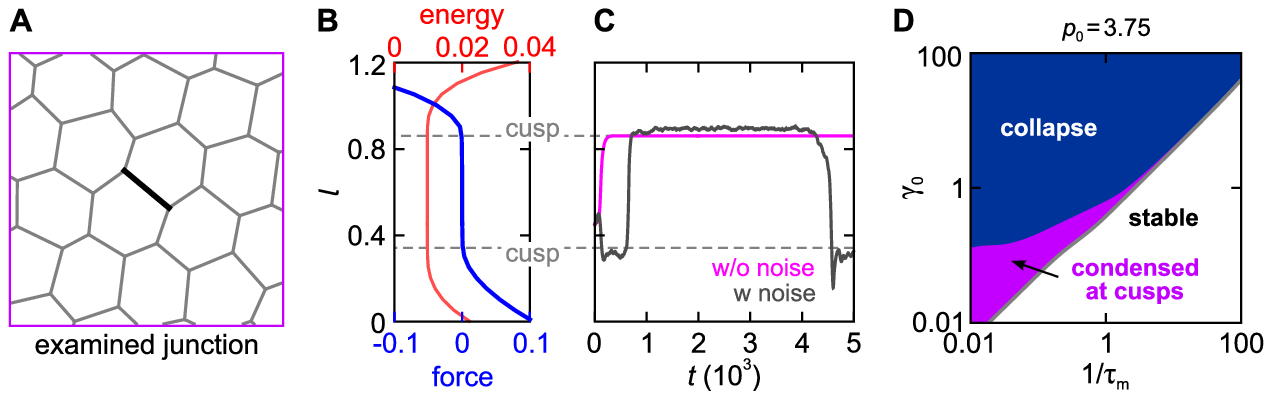}
		\caption{(A)~A close up of the examined junction. 
		(B,~C)~Energy landscape and restoring force (panel B) and length dynamics (panel C) for the examined active junction shown in panel A. Gray dashed lines denote locations of the cusps for the examined junction.By our definition, positions of the cusps correspond to $l$-values where the force $\left | f\right |$ drops below $10^{-3}$. In panels A-C, $\left (1/\tau_m,p_0,\gamma_0\right )=(0.025,3.75,0.030)$. (D)~Phase diagram in the $(1/\tau_m,\gamma_0)$ plane at $p_0=3.75$ for the examined junction.}\vspace{0.5cm}
	\end{figure}
	
	\begin{figure}[h]
		\centering
		\includegraphics{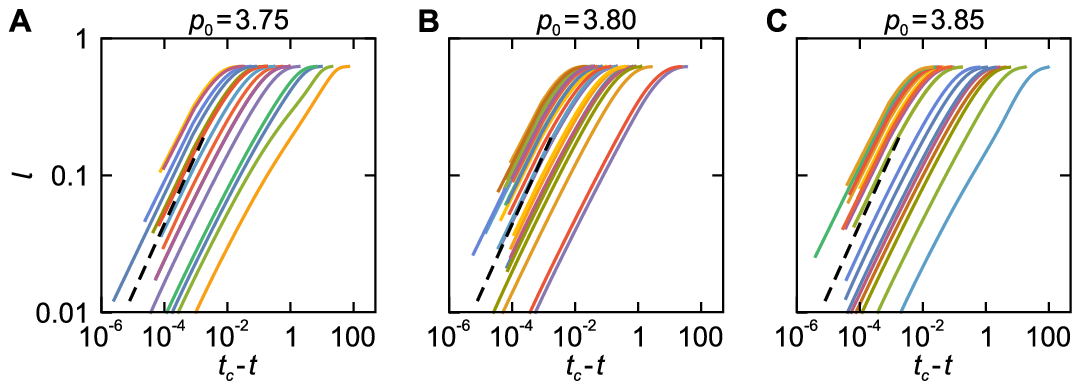}
		\caption{Junction length $l$ versus time $t_c-t$ (where $t_c$ is the time point of collapse) for active junctions in disordered tissues at $p_0=3.75,\>3.8,$ and $3.85$ (panels A, B, and C, respectively). Each panel shows trajectories for about 20 pairs of parameters $1/\tau_m$ and $\gamma_0$, chosen randomly from the regime of junction collapse. Dashed lines show $l(t)\sim(t_c-t)^{1/2}$.}\vspace{0.5cm}
	\end{figure}
	
	\begin{figure}[h]
		\centering
		\includegraphics{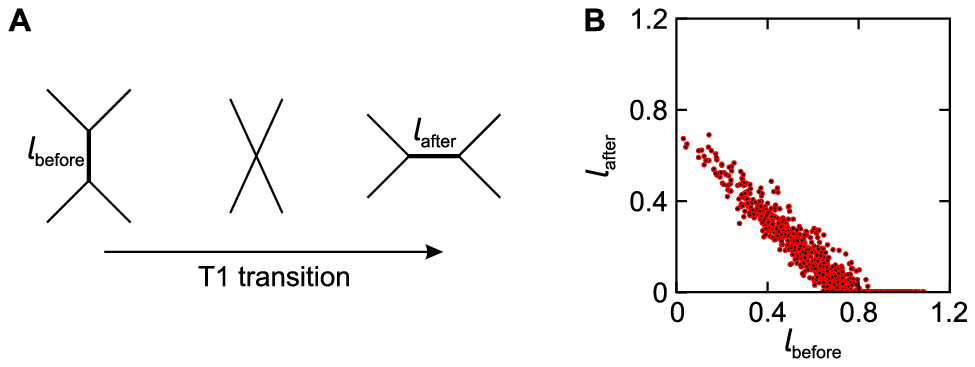}
		\caption{(A)~Schematic of a T1 transition showing the junction rest length before the transition, $l_{\rm before}$ and after the transition $l_{\rm after}$. (B)~The relation between the before- and after-T1 junction rest lengths at $p_0=3.65$. The results show that T1 transitions on average increase the rest lengths of initially short junctions and decrease the rest lengths of initially long junctions. In active junction networks, T1 transitions thus transform short (unstable) junctions into long (stable) junctions, which drives the overall tissue ordering.	}\vspace{0.5cm}
	\end{figure}
	
	\begin{figure}[h]
		\centering
		\includegraphics{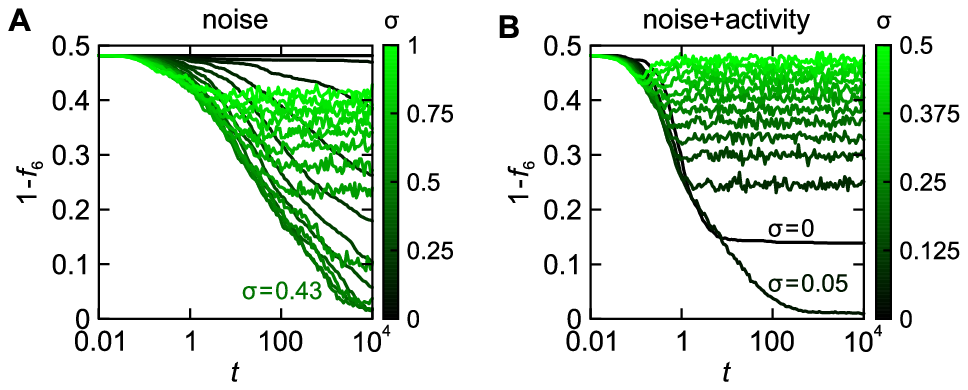}
		\caption{(A)~Fraction of non-hexagonal cells $1-f_6$ versus time $t$ in a model, where active-tension dynamics is described by a pure Ornstein-Uhlenbeck process. (B)~Fraction of non-hexagonal cells $1-f_6$ versus time $t$ in a model, where in addition to the Ornstein-Uhlenbeck process, the active-tension dynamics contains the term describing the positive feedback between junction contractions and generation of active tension (second term in Eq.~(1) of the main text). The black-to-green color scheme encodes the magnitude of tension fluctuations $\sigma$. In both panels, $1/\tau_m=19.05$ and $p_0=3.65$; additionally, in panel~B,~$\gamma_0=3.02$.}\vspace{0.5cm}
	\end{figure}
	